\documentclass[apj,iop,revtex4]{emulateapj}
\usepackage{natbib}
\usepackage{hyperref}
\usepackage{color}
\usepackage{xcolor}
\usepackage{amsmath}
\usepackage{graphicx,times}
\usepackage{natbib}
\usepackage{epstopdf}
\usepackage{txfonts}
\newcommand{\mytilde}{\raise.19ex\hbox{$\scriptstyle\sim$}}

\newcommand{\Rmnum}[1]{\expandafter\@slowromancap\romannumeral #1@}

\shorttitle{A Head-on Major Merger in the Nearby NGC 6338 Group}
\shortauthors{Y. Wang et al. 2019}

\begin{document}

\title{Revealing A Head-on Major Merger in the Nearby NGC 6338 Group with {\it Chandra} and VLA observations}

\author{
Yu Wang\altaffilmark{1,\dag}, Fuyao Lui\altaffilmark{1}, Zhiqiang Shen\altaffilmark{2}, \\
Jingying Wang\altaffilmark{3}, Dan Hu\altaffilmark{4} and Hai-Guang Xu\altaffilmark{4,5,6}
}

\altaffiltext{1}{School of Mathematics, Physics and statistics, Shanghai University of Engineering Science, 333 Longteng Road, Shanghai 201620, China; Email: wang\_yu@fudan.edu.cn}
\altaffiltext{2}{Shanghai Astronomical Observatory, Chinese Academy of Sciences, 80 Nandan Road, Shanghai 200030, China}
\altaffiltext{3}{Department of Physics and Astronomy, University of the Western Cape, Cape Town 7535, South Africa}
\altaffiltext{4}{School of Physics and Astronomy, Shanghai Jiao Tong University, 800 Dongchuan Road, Shanghai 200240, China }
\altaffiltext{5}{Tsung-Dao Lee Institute, Shanghai Jiao Tong University, 800 Dongchuan Road, Shanghai 200240, China}
\altaffiltext{5}{IFSA Collaborative Innovation Center, Shanghai Jiao Tong University, 800 Dongchuan Road, Shanghai 200240, China}

\begin{abstract}
By analyzing the {\it Chandra} archival data of the nearby NGC 6338 galaxy group,
we identify two X-ray bright clumps (N-clump and S-clump) within the central $100h_{73}^{-1}$ kpc,
and detect an arc-like X-ray brightness discontinuity at the south boundary of the N-clump,
which is defined as a cold front with a gas flow Mach number of $\mathcal M<0.8$.
Furthermore, at the north-east boundary of the S-clump (dominated by galaxy NGC 6338)
another X-ray edge is detected that corresponds to a weaker cold front.
Therefore, the two clumps are approaching each other approximately from opposite directions,
and the group is undergoing a head-on collision that is in a stage of pre-core passage.
This merger scenario is also supported by the study of the line-of-sight velocity distribution of the group member galaxies.
The merger mass ratio is about $1:1.8$ as estimated from
the central gas temperature of the two clumps, which suggests the merger is most likely to be a major merger.
We also analyze the VLA 1.4 and 4.9 GHz radio data,
but we do not detect any extended radio emission that is associated with the merger.
\end{abstract}

\section{Introduction}
In the framework of hierarchical structure formation,
galaxy clusters grow in size by merging with subunits,
releasing as much as $10^{64}$ erg of kinetic energy
as thermal energy by driving shocks \citep{sara2002},
with each merger event typically lasting for about $2-5$ Gyr
\cite[e.g.,][]{roet1997,asca2006}.
It is expected that major merger processes
can generate remarkable hydrodynamic substructures in the intracluster medium (ICM),
such as shocks and cold fronts that show arc-shaped or edge-like morphologies,
corresponding to gas density and temperature jumps
\citep[e.g.,][and references therein]{mark2007}.
These substructures can be used to determine the kinematics of the
merger and to study the conditions and transport processes in the
ICM, including electron-ion equilibrium and thermal conduction
\citep[e.g.,][]{mark2006}.

It is likely that a fraction of the shock energy can be converted
into the acceleration of relativistic particles
\citep[e.g.,][]{blan1987},
and in cluster mergers this process could produce synchrotron radio emission
\citep[e.g.,][]{fere2012,brun2014}.
Giant radio halos and relics have been observed in galaxy clusters
\citep[e.g.,][]{fere2008},
and recent years, many more radio halos and/or relics in lower frequencies
have been detected in relaxed clusters, even in poor clusters
\citep[e.g.,][]{fere2012},
some of which have been explained by gas sloshing in the core
\citep[e.g.,][]{zuho2013}.
However, as radio emitting electrons have short radiative lifetimes ($\sim 10^{7}-10^{8}$ yr),
it is difficult to explain the Mpc size of extended radio halos
\citep[for a review see][]{brun2014}.
And there exists some dynamically disturbed clusters that do not show any evidence of radio halo,
such as the well-known merging clusters A119 \citep{giov2000}.
Also, \citet{cass2010} found four such clusters: A141,
A781 (that has been found to host a radio halo in \citet{govo2011}, A2631 and MACS J2228.5+2036.
On the scale of galaxy group, some possible signals for extend radio sources
related to the intergalactic magnetic field have been found
\citep[e.g.,][]{niki2017},
while the evidence for radio relics or halos in the groups is not yet conclusive.
The above indicates that the formation mechanism of radio halos is currently not fully understood.

As we know, a large fraction of the baryons in the nearby Universe resides in galaxy groups
with X-ray luminosities $\sim 10^{41} - 10^{43}$ erg s$^{-1}$ and gas temperature $\sim 0.3-2$ keV
\citep[e.g.,][]{ponm1993,mulc2000}.
Galaxy groups are much more representative gravitational systems than rarer rich clusters of galaxies
\citep[e.g.,][]{gell1983,tull1987}
and important for understanding the gravitational and thermal evolution of most of the matter in the Universe.
The effects of non-gravitational heating, such as AGN feedback and merger shocks,
are expected to be more significant in lower mass systems
as the energy input from these sources is comparable to the binding energy of the group
\citep[e.g.,][]{ponm1996,ponm1999,hels2000}.
Major and minor group mergers, and their subsequent relaxation,
govern the formation of the largest scale structures
and can have a considerable impact on their constituent galaxies.
However, studies of galaxy group mergers have been limited due to their faint X-ray emission and low galaxy densities.
And there are not many research works focused on group mergers
\citep[e.g.,][]{kraf2006,kraf2011,mach2010,mach2011,russ2014,sche2017}
when compared with a large number of studies on galaxy cluster mergers.

In this paper, we present a nearly head-on merger
discovered in the {\it Chandra} observation of the nearby NGC 6338 galaxy group \citep[$z=0.02824$;][]{wegn1999}.
The brightest galaxy NGC 6338 (cD, S0, $M_{B}=13.6$; $z=0.02743$) and
the member galaxy 2MASX J17152326+5725585
\citep[hereafter 2MASX J1715; E0, $M_{B}=15.4$; $z=0.03212$;][]{smit2004}
present some evidence of galaxy-galaxy interaction in the optical band,
as shown on the SDSS image (Fig. $1a$).
Both NGC 6338 and 2MASX J1715 (located north about $1.3^{\prime}$ from NGC 6338) are elongated in the north-south direction;
In particular, the latter contains a double nucleus
\citep[][Fig. $1a$]{berl2006},
and has an asymmetrical optical morphology that are possibly due to tidal forces.
In the VLA NVSS map
\citep[][Fig. $1a$]{cond1998},
a point-like radio source is shown at the center of galaxy NGC 6338,
as reported by \citet{dong2010} and \citet{pand2012}.
We also find an extended radio emission located in the south-east side of the group,
which infers that the group has a radio relic.

In \S2, we describe the {\it Chandra} and VLA observations, and data reductions.
In \S3, we present the X-ray image and VLA radio map.
In \S4, we analyze the velocity distribution of the identified member galaxies.
In \S5, we discuss and summarize our results.
Throughout the paper, we adopt the cosmological parameters
$H_{0} = 73$ km s$^{-1}$ Mpc$^{-1}$,
$\Omega_{\rm b} = 0.044$,
$\Omega_{\rm M} = 0.27$,
and
$\Omega_{\Lambda} = 0.73$.
Unless stated otherwise, the quoted errors stand for 90$\%$
confidence limits.

\section{OBSERVATIONS AND DATA REDUCTIONS}
\subsection{\it Chandra}

The NGC 6338 galaxy group was observed by {\it Chandra}
on September 17-18, 2003 (47.94 ks, ObsID 4194)
with chips 0, 1, 2, 3, 6 and 7 of the Advanced CCD Imaging Spectrometer (ACIS) operating in VFAINT mode.
We use the {\it Chandra} data analysis package CIAO version 4.4 and CALDB version 4.5.1 to process the data,
by starting with the level-1 raw event files
in order to apply the latest corrections for the charge transfer inefficiency (CTI)
to improve the energy resolution of the CCDs
and to remove most of the effects of the apparent gain shift.
We keep events with {\it ASCA} grades 0, 2, 3, 4, and 6, and removed all the bad pixels, bad columns,
and columns adjacent to bad columns and node boundaries.
We examine the $0.3-12.0$ keV lightcurves extracted from the background regions defined on the four ACIS-I chips,
and find that there are almost no strong background flares
that increased the background count rate to $>115\%$ of the mean quiescent value.
The obtained net exposure is 47.13 ks.

For the spectral analysis,
we extract the {\it Chandra} spectra in the 0.7--7.0 keV band.
Background spectra are extracted from the {\it Chandra} blank-sky
fields; a cross-check based on the use of local background yields
essentially the same result.

\subsection{VLA}
We use the radio data of the NGC 6338 group observed with the Very Large Array (VLA) on
1998 July 28 at the frequency of 1.4 GHz (L-band) in B-array (AP3690)
and on 1997 June 27 at the frequency of 4.9 GHz (C-band) in C-array (AE0110).
The observations were carried out with 50 MHz bandwidth
for the total integration time of 950 seconds and 630 seconds for L-band and C-band, respectively.
In these observations 3C 48 is used as the primary flux density calibrator,
while 3C 343 and 1739+522 are used to determine the complex antenna gains, respectively.
The NRAO achieved data is analyzed in AIPS\footnote{https://www.aips.nrao.edu/index.shtml} (version 31DEC16) using the standard procedures.
Self-calibration is applied to remove residual phase variations.
The final images are produced by AIPS task IMAGR.

\section{RESULTS}
\subsection{X-ray Image}
In Figure $1b$, we show the ACIS-I image of the central
200 kpc $\times$ 200 kpc ($5.8^{\prime}\times5.8^{\prime}$)
region of the NGC 6338 group in the 0.3$-$5.0 keV band,
which has been corrected for exposure but not for background.
Two X-ray bright clumps are clearly visible in the center region
with a projected separation of about
$48h_{73}^{-1}$ kpc ($83^{\prime\prime}$; 1 arcsec=0.575 $h_{73}^{-1}$ kpc at the redshift of $z=0.02824$).
The X-ray peak of the south clump (hereafter S-clump)
is consistent with the optical centroid of NGC 6338 within $3^{\prime\prime}$,
and the X-ray peak of the north clump (hereafter N-clump) is found
at $2.8h_{73}^{-1}$ kpc ($5^{\prime\prime}$) north of member galaxy 2MASX J1715.
In Figure $1b$,
two arc-shaped edges can be seen at the south boundary of N-clump (hereafter N-edge)
and the north-east boundary of S-clump (hereafter S-edge);
and two X-ray stripped tails also are found located roughly north of N-clump
and south-west of S-clump, respectively.
These above findings indicate that two clumps are approaching each other
approximately from opposite directions,
and the group is undergoing a nearly head-on merger.

To study the edges in a quantitative way,
we extract the exposure-corrected X-ray surface brightness profiles
(SBPs; Fig. $2a$ and $2b$) in the $0.3-2.0$ keV band using two sets of semi-annuli,
which are defined in two semi-circles as shown in Figure $1b$.
We find that both profiles show clear surface brightness discontinuities at the edges;
across the N-edge,
the surface brightness increases inward by a factor of 2.1 within $3.6h_{73}^{-1}$ kpc ($6.3^{\prime\prime}$),
and across the S-edge,
it increases by a factor of 1.6 within $6.4h_{73}^{-1}$ kpc ($11.1^{\prime\prime}$).
Surface brightness discontinuities appear usually
due to the jumps in gas density accompanied with temperature jumps,
and often indicate the existence of shocks as in 
1E 0657$-$56 \citep{mark2002} and Abell 520 \citep{mark2005},
or cold fronts as in Abell 1795 \citep{mark2001} and Abell 168 \citep{hall2004}.

\subsection{Temperature and Metal Abundance Distributions}
In order to investigate the thermal properties of the head-on merging group,
we obtain the two-dimensional gas temperature and metal abundance distributions (Table 1)
by extracting spectra from the regions including
the semi-annuli of two clumps
(North semi-annuli of N-clump, hereafter NN; South semi-annuli of N-clump, SN;
North-east semi-annuli of S-clump, NS; and South-west semi-annuli of S-clump, SS) and Box (a--l) as shown in Figure 3.
We fit each spectrum with an absorbed APEC model coded in the XSPEC v12.4.0 software
by fixing the redshift and absorption to $z=0.02824$ and the Galactic value
$N_{\rm H} = 2.23 \times 10^{20}$ cm$^{-2}$ 
\citep[LAB Survey of Galactic HI,][]{kalb2005}, respectively.
Allowing the redshift to vary does not improve the fits.
Except for the spectrum extracted in the north middle semi-annulus of S-clump (Region NS2),
no more than Galactic absorption is needed.

In the central region, each of two clumps has a cool core
with the gas temperature of $1.0-1.3$ keV for N-clump and $1.3-1.5$ keV for S-clump.
The gas temperatures between two clumps (Region SN2 and NS3)
are up to $3.9$ keV, which are significantly higher than those of Regions NN2 and SS3 ($2.0-2.6$ keV).
With the two clumps approaching each other, the gas between two clumps is compressed,
and it is most likely heated by shocks generated in the merger.
In the surroundings, the gas temperatures of Boxes (a), (e), (f), and (i) with the value of $3.2-4.0$ keV
are significantly higher than those of Boxes (c) (d), (g), (h), (k), and (l) with the value of $1.9-2.5$ keV.
The above four high-temperature boxes are on either sides of the central merger region,
the gas of which is probably the hot gas expelled from the collision axis areas in the merger process,
as shown in simulation works \citep[e.g.,][]{rick2001}.

The stripped gas of the S-clump [Boxes (g), (k) and (l)] has a higher
metal abundance in a range of $0.5 - 1.5$ Z$_{\odot}$ (Table 1),
comparing with the average metal abundance of $0.25^{+0.08}_{-0.05}$ Z$_{\odot}$ 
\citep{rasm2007}.
In the tail of N-clump [Boxes (c) and (d)] no enhanced metal abundance is found.

\subsection{Two Cold Fronts}

We show the gas temperature profiles across the N-edge and S-edge in Figure 2($c-d$).
After correction for projection effects (Table 1), we find that the gas temperatures
outside both of the edges ($4.0^{+1.0}_{-0.7}$ keV and $3.6^{+0.5}_{-0.4}$ keV ) are significantly higher than
those inside ($1.1\pm0.1$ keV and $1.9\pm0.1$ keV ) at $90\%$ confidence level,
which confirm the existence of two temperature jumps at the N-edge and S-edge, respectively.
Note that the temperature jumps cannot be smeared out by abundance variations allowed by the data;
to show this, in Figure 4 we plot the two-dimensional fit-statistic contours
of temperature and abundance at 68\%, 90\%, and 99\% confidence levels
inside and outside N-edge and S-edge, respectively,
all obtained in the above deprojected fittings.

To estimate the gas flow of the two clumps,
we attempt to fit each of the exposure-corrected SBPs with two density models, respectively,
by applying the best-fit deprojected spectral parameters (Table 1).
The first density model (model A) is composed of two $\beta$ components as
\begin{equation}
n_{g}(R) = \left\{ n^{2}_{g,1} \left[ 1 + (R/R_{\rm c1})^{2} \right] ^{-3\beta_{1}}
           + n^{2}_{g,2} \left[ 1 + (R/R_{\rm c2})^{2} \right] ^{-3\beta_{2}}\right\}^{1/2},
\end{equation}
where $R$ is the 3D radius, $R_{\rm c}$ is the core radius, and $\beta$ is the slope.
The second density model (model B)
is composed of one truncated power-law component and one $\beta$ component,
\begin{equation}
n_{g}(R) = \left \{ \begin{array}{ll}
                     n_{g,1}(R/R_{\rm cut})^{-\alpha} & R<R_{\rm cut} \\
                     n_{g,2} \left[ 1 + (R/R_{\rm c})^{2} \right] ^{-3\beta/2}& R \geq R_{\rm cut}
                    \end{array}
           \right. ,
\end{equation}
where $R_{\rm cut}$ is the truncation radius to be determined in the fittings.
For each of the SBPs, an acceptable fit is obtained with model B only (Table 2; Fig. 2),
and we show the gas density inside and outside of these two edges in Table 3.
The density jump is in a factor of $2.0^{+0.4}_{-0.3}$ across the N-edge, and $1.4\pm0.1$ across the S-edge.

Following the method of \citet{vikh2001},
to estimate the Mach number of gas flow
we need to investigate the gas pressure in both the stagnation point and the undisturbed free stream.
The gas pressure at the stagnation point must be equal to that inside the edge,
which can be well determined by the X-ray gas density and temperature of the N-clump and S-clump (Table 1 and Table 3).
Because the gas between the two clumps (Region SN2 and NS3; Table 1) is significantly heated by
the merger as shown in \S3.2, it cannot be thought of as the gas temperature of free stream.
Thus, we assume the undisturbed gas in the north semi-annulus of N-clump (Region NN2)
and the south-west semi-annulus of S-clump (Region SS3) as the free stream.
Then, we calculate the average thermal gas pressure inside of the edge and in the corresponding free stream
as $P_{\rm in}$ and $P_{\rm 0}$,
and find that the pressure ratio is
$P_{\rm in}/P_{\rm 0}=1.1^{+0.5}_{-0.3}$ for N-edge and $1.0\pm0.3$ for S-edge (Table 3), respectively,
which indicates that the pressure equilibrium has been established across the two edges.
Also, from Bernoulli's equation \citep{land1959}
we estimate that the Mach number of the gas flow is $\mathcal M_{\rm N-edge}<0.8$
(corresponding to a velocity of $580$ km s$^{-1}$) for the N-clump.
Because the above is a rough estimate and no shock-like edge is found in the X-ray image,
it is believed that the N-clump moves toward the south in a subsonic flow,
which is consistent with that found by \citet{dupk2013}.
For the S-edge, the Mach number is $\mathcal M_{\rm S-edge}<0.6$,
corresponding to a velocity less than $430$ km s$^{-1}$.
Based on the above results, we conclude that
the N-clump comes from north to merge with the S-clump, and a cold front is observed at its south boundary.
At the same time, the S-clump is moving north-east at a slower speed,
and correspondingly, the other cold front is formed at the north-east boundary of S-clump.
Therefore, the head-on merger is in a stage of the pre-core passage.

\subsection{Radio Properties}
In Figure 5 ($a$) and ($b$), we show the VLA 1.4 GHz and 4.9 GHz contours of the central
200 kpc $\times$ 200 kpc ($5.8^{\prime}\times5.8^{\prime}$) region of NGC 6338 group.
Their synthesized beams and noise levels are
$\theta_{1.4}=3.3^{\prime\prime}\times5.6^{\prime\prime}$ and $\sigma_{1.4} = 0.01$ mJy/beam,
$\theta_{4.9}=3.4^{\prime\prime}\times5.1^{\prime\prime}$ and $\sigma_{4.9} = 0.03$ mJy/beam, respectively.
In Figure 6, we plot the 1.4 GHz contours on the exposure-corrected X-ray image.
We find no extended radio emission directly associated with the merger of NGC 6338 group in 1.4 and 4.9 GHz maps.

In the 1.4 GHz map, the radio source in the center of galaxy NGC 6338 is resolved,
and the radio emission displays an X-shaped appearance.
The radio lobes, extending to the north-east and south-west directions,
are associated with two possible X-ray cavities (Fig. 6),
which \citet{pand2012} found by analyzing {\it Chandra} X-ray data.
These indicate that these two radio lobes were inflated by jets
originating from the central AGN activity.
The radio substructures, extending to the north-west and south-east
with the size scale of 8.6$h_{73}^{-1}$ kpc (15$^{\prime\prime}$) and 7.7$h_{73}^{-1}$ kpc (13$^{\prime\prime}$) over $10\sigma$, respectively,
are roughly perpendicular to the older radio lobes,
and they are coincident with two X-ray filaments \citep[Fig. 6;][]{pand2012} 
and HST H$\alpha$ filaments \citep{mart2004}.
\citet{pand2012} also detected HST I-band filaments (usually associated with the interstellar dust) 
in the center of galaxy NGC 6338,
which extends up to about 3.6 kpc in the south-east direction.
The spatial correspondence of radio, X-ray, H$\alpha$ and optical filaments provides a strong evidence
that these features has been inflated by jets from the central AGN in its latest outburst.
In the VLA 4.9 GHz map, the radio source is not resolved (Fig. $5b$).

To the south-east of the NGC 6338 group, we find a radio galaxy with two lobes in both of 1.4 and 4.9 GHz maps,
which possibly correspond to a background AGN.
In the central region of the two lobes,
we find an X-ray point-like source that is
likely associated with the galaxy SSTSL2 J171538.78+572325.8 within $3^{\prime\prime}$, as shown in Figure 6.

\section{Velocity Distribution}
To investigate the merging status of the NGC 6338 group,
we select galaxies from the NASA/IPAC Extragalactic Database (NED)
within a radius of $31^{\prime}$ ($\sim$1 Mpc) centered at the weighted centroid of the group galaxy distribution
\citep[RA=17h15m24s Dec=+57d24m39.6s;][]{pear2015}.
To including possible substructures a velocity cut of $\pm2,000$ km s$^{-1}$ centered on the group redshift (z=0.02824) is applied.
Then, 82 galaxies are identified to belong to the group,
in which the two nuclei of 2MASX J1715 are defined as two galaxies.

First, we plot the line-of-sight velocity distribution of the member galaxies in Figure 7($a$),
which shows that a high-velocity plateau is located at about $9,800$ km s$^{-1}$.
We fit the observed distribution with a single Gaussian profile,
and then we calculate the Kolmogorov$-$Smirnov statistic for the observed distribution
against the best-fit Gaussian model ($\chi^{2}/{\rm d.o.f.}=63.8/15$),
which shows that the observed distribution has a probability of $<20\%$ of being Gaussian.
We attempt to fit the observed distribution with a two-component Gaussian model.
The best-fit ($\chi^{2}/{\rm d.o.f.}=23.7/12$) gives an average velocity of
$<v_{1}>=8,600\pm40$ km s$^{-1}$ and a corresponding variance of $\sigma_{v,1}=430^{+40}_{-30}$ km s$^{-1}$
for the main Gaussian component, and
$<v_{2}>=9,690^{+60}_{-50}$ km s$^{-1}$ and $\sigma_{v,2}=160^{+50}_{-40}$ km s$^{-1}$
for the high-velocity plateau.
By applying the F-test and the Kaye's Mixture Model 
\citep[KMM;][]{mcla1988,ashm1994} test,
the latter of which is based on a maximum likelihood algorithm,
we find that the second Gaussian component is required at the 99\% confidence level and preferred
at a significant probability of 90\%, respectively.

Following the method of \citet{wang2010},
to investigate whether the galaxies in the high-velocity plateau form a real substructure,
we divide the identified galaxies into two sub-groups:
one sub-group with a low-velocity of $7,500-9,500$ km s$^{-1}$)
and the other with a high-velocity of $9,500-10,600$ km s$^{-1}$),
which consist of 70 and 12 galaxies, respectively.
According to the best-fit two-Gaussian model,
these sub-groups roughly corresponds to two Gaussian components, respectively,
with up to about two of the galaxies in the high-velocity sub-group coming from the main Gaussian component.
As shown in Figure 7($b$), we find that the galaxies belonging to the
high-velocity sub-group are distributed mostly in the north-east part,
which includes the member galaxies 2MASX J1715 and NGC 6345 (S/S0, $M_{B}=14.8$; $z=0.03449$).
The galaxies in the low-velocity sub-group that is dominated by galaxy NGC 6338,
on the other hand, are scattered symmetrically in the field.
These results suggest that the galaxy velocity separation
has a dynamical nature and the group is undergoing a merger.

\section{DISCUSSION AND SUMMARY}
Based on our X-ray analysis, we find that
in the central 100$h_{73}^{-1}$ kpc of NGC 6338 group
an X-ray brightness discontinuity is detected
at the south boundary of the N-clump (associated with the high-velocity sub-group),
and at the north-east boundary of the S-clump (related to the low-velocity sub-group)
the other X-ray edge is also found,
each of which is defined as a cold front.
Therefore, the group is undergoing a head-on collision that is in the pre-core passage stage.

As shown in Figure 1b and Table 1, both of the sub-groups have a cool core that is not significantly affected by the merger.
Because both N-clump and S-clump are gas rich, we can use their X-ray properties to simply estimate the mass ratio of two sub-groups.
We get the spectrum of N-clump extracting from both Regions NN1 and SN1, and that of S-clump from all the Region NS1, NS2, SS1, and SS2.
Then we fit each spectrum with the absorbed APEC model as used in \S3.2.
The obtained gas temperatures are $T_{N}=1.16\pm{0.07}$ keV and $T_{S}=1.66\pm{0.04}$ keV for N-clump and S-clump, respectively.
According to the study of \citet{huds2010}, cool-core clusters have a systematic central temperature drop,
as $T_{c}\propto 0.4T_{vir}$, where $T_{c}$ is the cool-core temperature and $T_{vir}$ is the virial temperature of the galaxy cluster.
Basing on the scale relation of $M_{vir}\propto T_{vir}^{1.65}$  for galaxy groups and clusters \citep{sun2009},
the merger mass ratio of two sub-groups is
R = $\frac{M_{N}}{M_{S}} \propto (\frac{T_{vir,N}}{T_{vir,S}})^{1.65} \propto (\frac{T_{N}}{T_{S}})^{1.65} = 1:1.8$,
which supports the merger is most likely to be a major merger.
Here, we do not use the velocity variance ratio of two sub-groups (\S4) to calculate the merger mass ratio,
because the number of member galaxies of the high-velocity sub-group is small (only 12 galaxies)
and the uncertainty of error will be large.

In \S3.3, the speed of N-clump is $< 1,010$ km s$^{-1}$ relative to S-clump roughly in the plane of sky.
And in \S4, the high-velocity subgroup has a velocity of
$\sim 1,090$ km s$^{-1}$ relative to the low-velocity subgroup along the line-of-sight direction.
According to the Pythagorean Theorem, the high-velocity sub-group (N-clump)
has a comprehensive velocity of $< 1,490$ km s$^{-1}$ to the low-velocity sub-group (S-clump).
Assuming the two clumps move along the projected distance $48h_{73}^{-1}$ kpc at their currents velocities,
they would totally collide with $>30$ million year.

Considering the relation between radio relic luminosity and system total mass \citep{dega2014},
and the total mass of NGC 6338 group as $M_{500}=9.0\pm0.5\times10^{13} {\rm M_{\odot}}$ 
\citep{pear2015},
the expected luminosity of one radio relic would be $2.0\times10^{22}$ Watt Hz$^{-1}$,
corresponding to the flux density of about 10 mJy at 1.4 GHz at the redshift of NGC 6338 group.
One radio relic with a flux of 1.0 mJy and a size scale of $10\times50$ kpc$^2$ ($17.4^{\prime\prime}\times87.0^{\prime\prime}$)
would be detected at the present flux sensitivity of 0.01 mJy/beam
($\theta_{1.4}=3.3^{\prime\prime}\times5.6^{\prime\prime}$; \S3.4).
However, no extended radio emission directly associated with the merger of NGC 6338 group is found (\S3.4).
It is possible that the group has a lower merger-caused radio luminosity than that the luminosity-mass relation predicts.
In addition, radio relics and halos are usually expected to form after core passage, while the merger in NGC 6338 group in still in the pre-core passage stage.
This could be another reason why no merger-caused radio emission is detected in the group.

In the center of galaxy NGC 6338, we find an X-shaped radio structure in the 1.4 GHz map,
which is spatially overlapped by two pairs of radio lobes possibly caused by two different AGN activities.
The two radio lobes in the north-east and south-west directions,
associated with two possible X-ray cavities (Fig. 6),
are inflated by jets in an AGN activity;
and the others in the north-west and south-east directions
originate from the center AGN in its latest outburst,
because they spatially correspond to X-ray jets, H$\alpha$ and optical filaments.
Employing a simple assumption that a radio bubble is launched from the nucleus and travels at
approximately the sound speed ${\it c}_{s}$, and the time it takes to rise to its projected
position is the sound crossing time
$t_{\it c} = R/{\it c}_{s}$,
where the sound speed is ${\it c}_{s} = \sqrt{\gamma kT/(\mu m_{H})} \approx 1100\sqrt{T/5{\rm keV}}$ km s$^{-1}$
with $\gamma \approx 5/3$, and $\mu \approx 0.62$.
The estimated inflated time is ~$1.5\times10^{7}$ year for the younger lobes.
This indicates that the AGN in the center of galaxy NGC 6338
changed the direction of its jets about $1.5\times10^{7}$ year ago,
which is usually explained by axis precession
\citep[e.g.,][]{ster2008,falc2010}.

\acknowledgments
{We thank the {\it Chandra} team for making data available via the High
Energy Astrophysics Science Archive Research Center (HEASARC).
This work was supported by the Ministry of Science and Technology of China (grant No. 2018YFA0404601),
the National Science Foundation of China (grant Nos. 11103057, 11433002, 11533004, 11621303 and 61371147).
}

\bibliographystyle{aasjournal.bst}

\begin{table}
\caption{Gas Temperature and Metal Abundance Distributions}
\centering
\begin{tabular}{lccccc}
\hline\hline
Region & Radius &$N_{H}$ &$T$ & $Z$ & $\chi^2/dof$ \\
&($h_{73}^{-1}$ kpc)&  $10^{20}$ cm$^{-2}$ &(keV)&(Z$_{\odot}$)& \\
\hline
\multicolumn{6}{c}{PROJECTED$\tablenotemark{a}$}\\
\hline
NN1& 0$-$8.0&fixed&  $1.00^{+0.06}_{-0.06}$&$0.21^{+0.11}_{-0.07}$&62.7/49\\
NN2&8.0$-$17.9&fixed&$1.97^{+0.19}_{-0.21}$&$0.70^{+0.45}_{-0.31}$&57.5/59\\
SN1& 0$-$8.0&fixed  &$1.27^{+0.08}_{-0.11}$&$0.16^{+0.09}_{-0.07}$&49.0/56\\
SN2&8.0$-$17.9&fixed&$3.92^{+1.03}_{-0.75}$&$0.51^{+0.71}_{-0.48}$&38.4/58\\
\hline
NS1&  0$-$8.5&fixed  &$1.49^{+0.06}_{-0.07}$&$0.84^{+0.28}_{-0.22}$&76.6/57\\
NS2&8.5$-$25.6&fixed &$2.24^{+0.11}_{-0.10}$&$1.21^{+0.27}_{-0.23}$&53.4/57\\
NS2&8.5$-$25.6&$10.93^{+3.70}_{-3.42}$&$2.01^{+0.10}_{-0.12}$&$0.90^{+0.24}_{-0.20}$&35.6/56\\
NS3&25.6$-$39.3&fixed&$3.90^{+0.66}_{-0.54}$&$0.51^{+0.44}_{-0.35}$&78.77/57\\
SS1&  0$-$8.5&fixed  &$1.27^{+0.03}_{-0.03}$&$0.73^{+0.22}_{-0.15}$&99.4/58\\
SS2&8.5$-$25.6&fixed &$1.91^{+0.11}_{-0.10}$&$1.40^{+0.65}_{-0.32}$&55.0/59\\
SS3&25.6$-$39.3&fixed&$2.63^{+0.42}_{-0.29}$&$0.74^{+0.61}_{-0.37}$&80.3/59\\
\hline
a& &fixed&$3.95^{+0.37}_{-0.37}$&$0.79^{+0.45}_{-0.36}$&42.1/59\\
b& &fixed&$2.87^{+0.54}_{-0.49}$&$0.26^{+0.43}_{-0.26}$&55.5/59\\
c& &fixed&$2.49^{+0.39}_{-0.29}$&$0.23^{+0.23}_{-0.20}$&66.2/59\\
d& &fixed&$1.91^{+0.14}_{-0.17}$&$0.33^{+0.17}_{-0.12}$&86.4/59\\
e& &fixed&$3.19^{+0.29}_{-0.29}$&$0.50^{+0.30}_{-0.24}$&68.7/59\\
f& &fixed&$3.92^{+0.70}_{-0.60}$&$0.49^{+0.58}_{-0.42}$&46.6/59\\
g& &fixed&$2.51^{+0.21}_{-0.22}$&$0.88^{+0.45}_{-0.31}$&52.2/59\\
h& &fixed&$2.45^{+0.27}_{-0.28}$&$0.41^{+0.30}_{-0.21}$&64.5/59\\
i& &fixed&$3.49^{+0.53}_{-0.40}$&$0.69^{+0.56}_{-0.42}$&57.1/59\\
j& &fixed&$2.95^{+0.41}_{-0.39}$&$0.33^{+0.33}_{-0.25}$&54.8/59\\
k& &fixed&$2.66^{+0.30}_{-0.21}$&$0.77^{+0.38}_{-0.27}$&56.3/57\\
l& &fixed&$2.30^{+0.31}_{-0.27}$&$0.92^{+0.57}_{-0.37}$&59.2/59\\
\hline
\multicolumn{6}{c}{DEPROJECTED\tablenotemark{b}}\\
\hline
SN1& 0$-$8.0&fixed&$1.06^{+0.13}_{-0.07}$&$0.19^{+0.28}_{-0.08}$&101.1/117\\
SN2&8.0$-$17.9&fixed&$4.04^{+0.99}_{-0.70}$&$0.66^{+0.69}_{-0.49}$&\\
\hline
NS1&  0$-$8.5&fixed&$1.35^{+0.06}_{-0.03}$&$0.80^{+0.27}_{-0.19}$&216.8/173\\
NS2&8.5$-$25.6&fixed as 10.93&$1.87^{+0.10}_{-0.11}$&$0.98^{+0.34}_{-0.25}$&\\
NS3&25.6$-$39.3&fixed&$3.56^{+0.54}_{-0.44}$&$0.53^{+0.42}_{-0.32}$&\\
\hline\hline
\end{tabular}
\tablenotetext{1}{An absorbed APEC model is used to fit the spectra extracted from the semi-annuli
centered on the X-ray peaks of N-clump (Region NN$1-2$ and SN$1-2$) and S-clump (Region NS$1-3$ and SN$1-3$), and
Box (a$-$l) as shown in Fig. 3.
The redshift and absorption are fixed to $z=0.02824$ and the Galactic value $N_{\rm H} = 2.23 \times 10^{20}$ cm$^{-2}$
(LAB Survey of Galactic HI;	Kalberla et al. 2005), respectively.}
\tablenotetext{2}{The PROJCT model in XSPEC v12.4.0 is used to deproject the spectra in the south semi-annuli of N-clump (Region SN$1-2$) and the north-east semi-annuli of S-clump (Region SN$1-3$), respectively.}
\end{table}

\begin{table}
\caption{Acceptable Fits to the Surface Brightness Profiles across the N-edge and S-edge\tablenotemark{a} }
\centering
\begin{tabular}{lccccccc}
\hline
&$n_{g,1}$&$\alpha$&$R_{\rm cut}$&$n_{g,2}$&$\beta$&$R_{\rm c}$&$\chi^2/dof$\\
&($10^{-3}$ cm$^{-3}$) &&($h_{73}^{-1}$ kpc)& ($10^{-3}$ cm$^{-3}$) &&($h_{73}^{-1}$ kpc)&\\
\hline
N-clump&22.5$\pm2.2$&0.47$^{+0.06}_{-0.07}$&8.1$^{+0.4}_{-0.2}$&18.7$^{+1.9}_{-1.7}$&0.70$^{+0.03}_{-0.02}$&12.3$\pm0.3$&2.28/3\\
S-clump&6.3$\pm0.4$&1.37$\pm0.03$&25.8$\pm0.2$&5.5$\pm0.7$&0.17$^{+0.08}_{-0.10}$&73.1$^{+24.4}_{-13.6}$&3.47/3\\
\hline
\end{tabular}
\tablenotetext{1}{
Two density models (Model A and B) are used to fit the SBPs across the N-edge and S-edge (Fig. 2),
which are extracted from the south semi-annuli of N-clump and the north-east semi-annuli of S-clump (Fig. 1), respectively.
Best-fit deprojected spectral parameters (Table 1) are adopted in the fittings to calculate gas emission.
For each of SBPs, the acceptable fit is obtained with model B only. We show them here.}
\end{table}

\begin{table}
\centering
\caption{Gas parameters across the N-edge and S-edge }
\begin{tabular}{lccc}
\hline
Region&$n_{g}$&$T$&$P=Tn_{g}$\\
&($10^{-4}$ cm$^{-3}$) &(keV)&($10^{-4}$ keV cm$^{-3}$)\\
\hline
Inside the N-edge &  $1.30\pm0.09$& $1.06^{+0.13}_{-0.07}$& $1.4^{+0.3}_{-0.2}$\\
Outside the N-edge & $0.64\pm0.06$& $1.97^{+0.19}_{-0.21}\tablenotemark{a}$ & $1.3^{+0.2}_{-0.3}$\\
Inside the S-edge &  $0.50\pm0.02$& $1.87^{+0.10}_{-0.11}$& $0.9^{+0.1}_{-0.1}$ \\
Outside the S-edge & $0.36\pm0.02$& $2.63^{+0.42}_{-0.29}\tablenotemark{b}$& $0.9^{+0.3}_{-0.1}$\\
\hline
\end{tabular}
\tablenotetext{1}{Gas temperature in Region NN2 (Table 1 and Fig. 3) to be assumed as that of the free stream for the N-edge.}
\tablenotetext{2}{Gas temperature in Region SS3 (Table 1 and Fig. 3) to be assumed as that of the free stream for the S-edge.}
\end{table}

\begin{figure}
\centering
\includegraphics[width=15.0cm,angle=0]{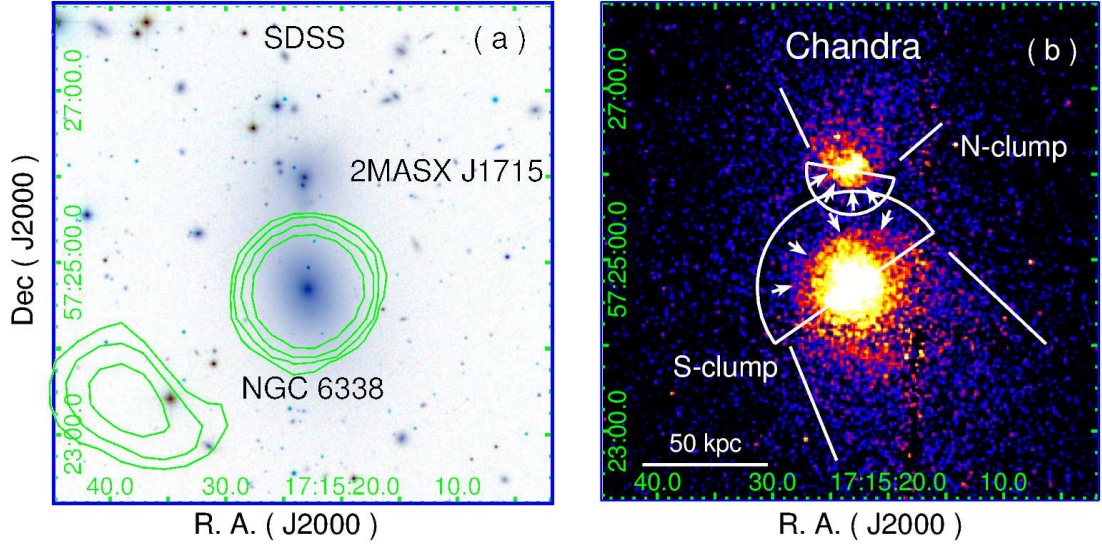}
\caption{
($a$) SDSS inverted-color image in central $200\times200$ kpc$^{2}$($5.8^{\prime}\times5.8^{\prime}$)
region for NGC 6338 group. The green contours present the NVSS flux density.
The beam is $45^{\prime\prime}\times45^{\prime\prime}$ and contour levels are 1.1, 2.0, 4.0, 8.0 mJy/beam.
($b$) Exposure-corrected and adaptively smoothed {\it Chandra}
ACIS-I image in the 0.3--5.0 keV band for the group in the same region as ($a$).
Two half-circles are overlaid, which are used to extract SBPs.
Two sets of arrows are used to show the positions of the N-edge (the south boundary of N-clump)
and S-edge (the north-east boundary of S-clump).
Two sets of lines at the north of N-clump and south-west of S-clump are used to show
the X-ray stripped tails for N-clump and S-clump, respectively.
}
\end{figure}

\begin{figure}
\centering
\includegraphics[width=12.0cm,angle=0]{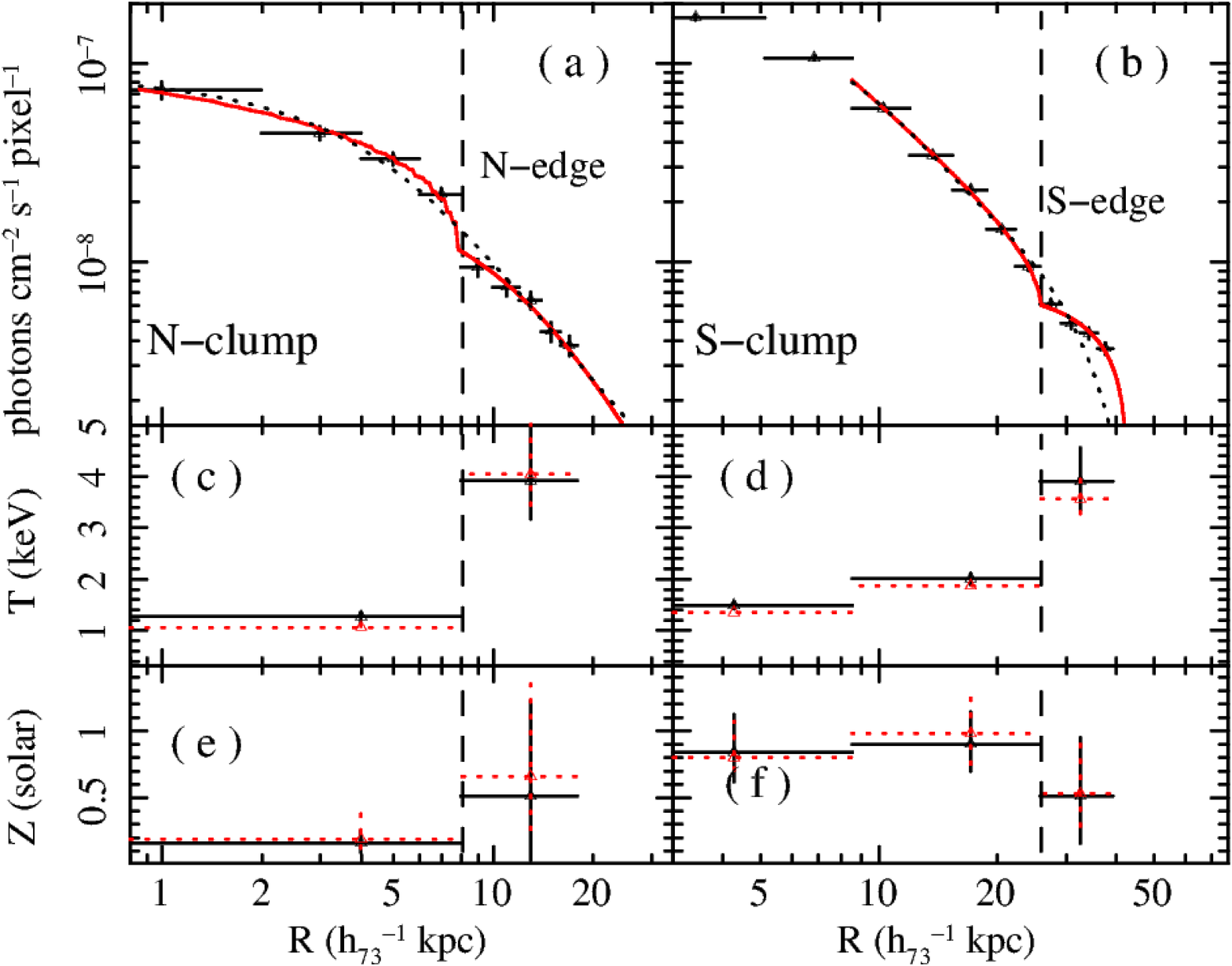}
\caption{
({\it up}) {\it Chandra} ACIS-I SBPs in the $0.3-2.0$ keV band for N-clump ($a$) and S-clump ($b$), respectively,
extracted from the south semi-annuli of the N-clump and the north-east semi-annuli of the S-clump as shown in Fig. 1.
Dotted (model A: two $\beta$ components) and solid (model B: one truncated power-law + $\beta$ components) lines show the best-fits
obtained with the two density models described in \S3.3, respectively.
({\it middle}) {\it Chandra} ACIS-I temperature profiles for N-clump ($c$) and S-clump ($d$),
extracted from the same region as SBPs.
({\it down}) {\it Chandra} ACIS-I metal abundance profiles for N-clump ($e$) and S-clump ($f$),
extracted from the same region as SBPs.
The dotted data in ($c-f$) present the deprojected spectral results.
All the values of gas temperature and metal abundance are listed in Table 1.
}
\end{figure}

\begin{figure}
\centering
\includegraphics[width=12.0cm,angle=0]{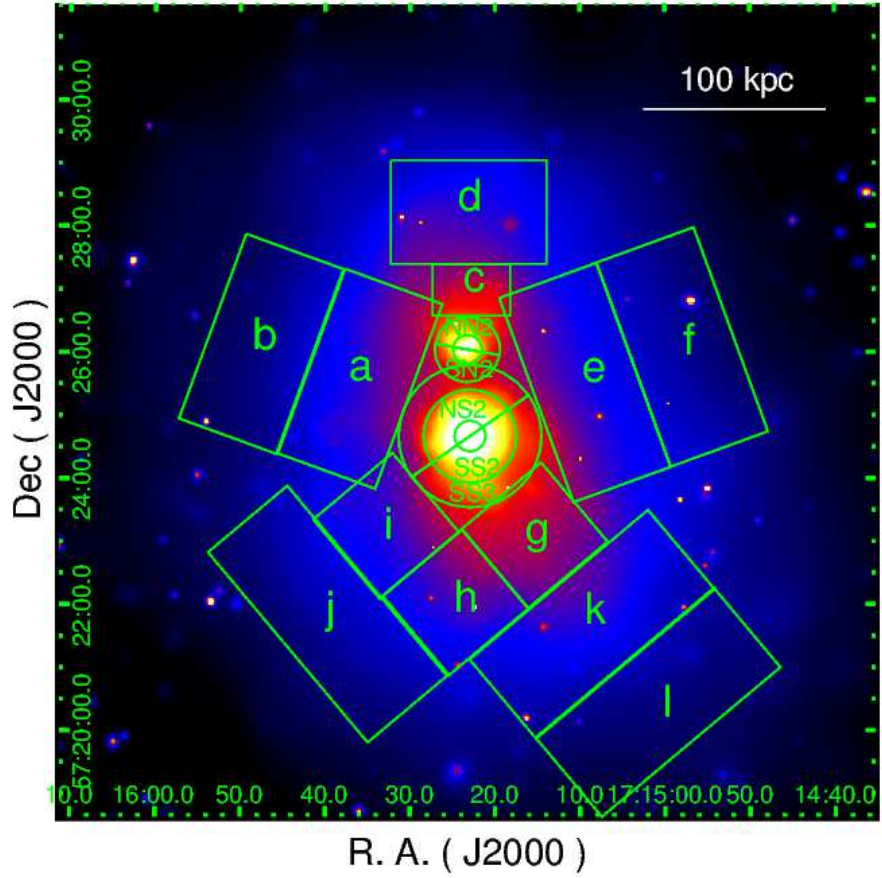}
\caption{
Exposure-corrected and background-subtracted {\it Chandra}
ACIS-I image in the central $450\times450$ kpc$^{2}$($13.1^{\prime}\times13.1^{\prime}$) region
for NGC 6338 group, overlaying the regions to extracted spectra as listed in Table 1.
}
\end{figure}

\begin{figure}
\centering
\includegraphics[width=15.0cm,angle=0]{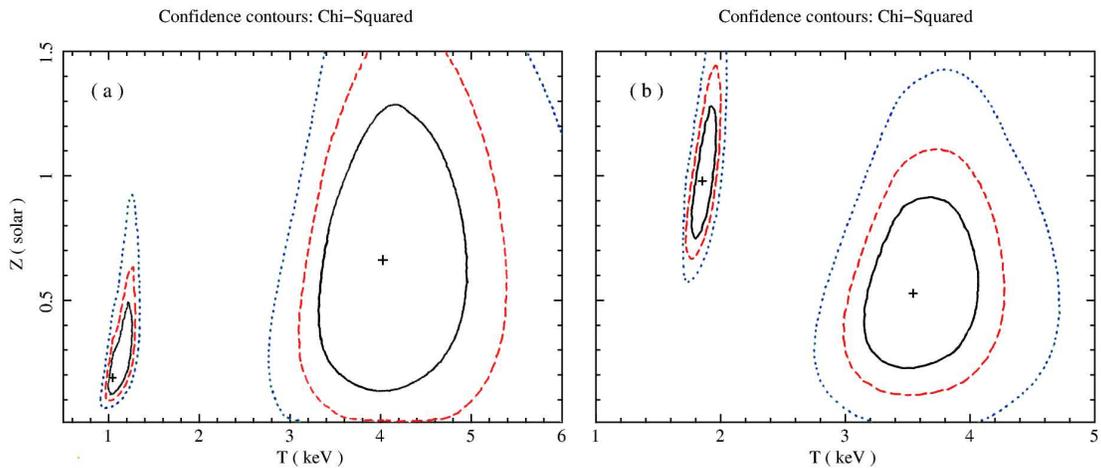}
\caption{
Fit-statistic contours at 68\% (solid), 90\% (dashed), and 99\% (dotted) confidence levels
for the gas temperature and metal abundance derived with the deprojected APEC model
in Region SN$1-2$
($a$) across the N-edge (Fig. $2c$ and $e$)
and Region NS$2-3$
($b$) across the S-edge (Fig. $2d$ and $f$), respectively.
}
\end{figure}

\begin{figure}
\centering
\includegraphics[width=15.0cm,angle=0]{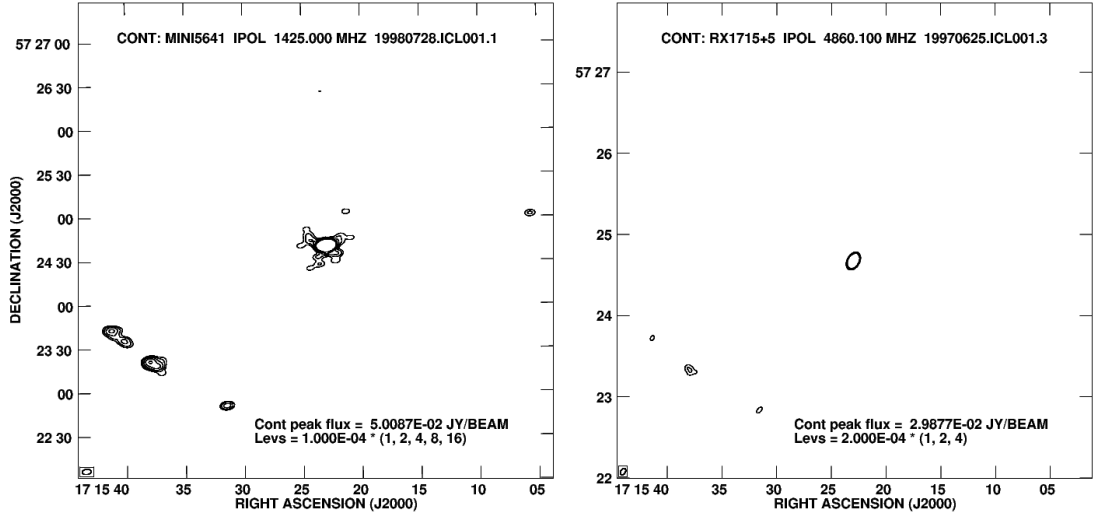}
\caption{
($a$) VLA 1.4 GHz contours
($\theta_{1.4}=3.3^{\prime\prime}\times5.6^{\prime\prime}$, $\sigma_{1.4} = 0.01$ mJy/beam,
and contour levels = 0.1, 0.2, 0.4, 0.8, 1.6 mJy/beam)
in central $200\times200$ kpc$^{2}$($5.8^{\prime}\times5.8^{\prime}$) region for NGC 6338 group.
($b$) VLA 4.9 GHz contours
($\theta_{4.9}=3.4^{\prime\prime}\times5.1^{\prime\prime}$, $\sigma_{4.9} = 0.03$ mJy/beam,
and contour levels = 0.2, 0.4, 0.8 mJy/beam)
in the same region as ($a$).
}
\end{figure}

\begin{figure}
\centering
\includegraphics[width=12.0cm,angle=0]{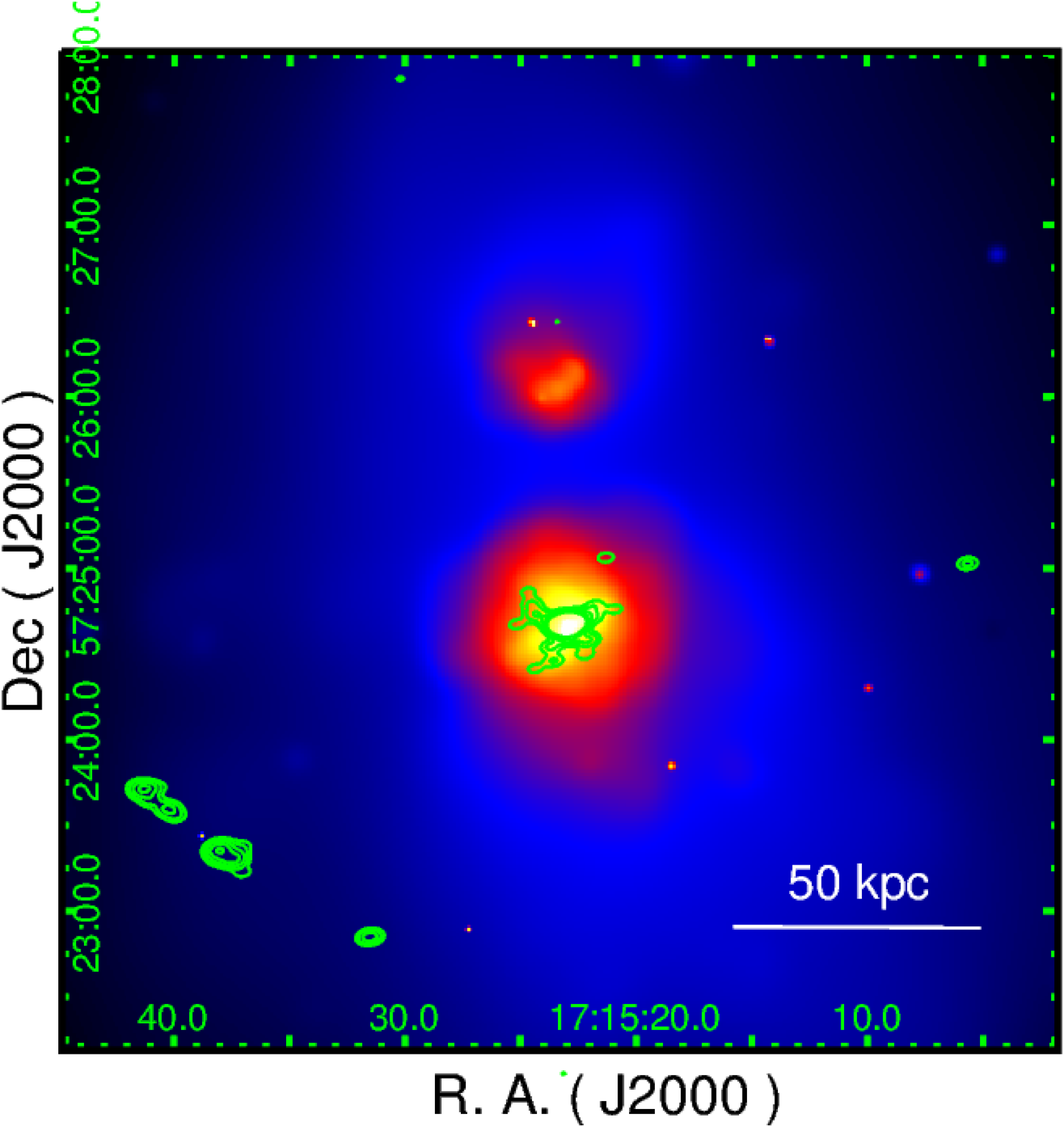}
\caption{
Exposure-corrected and background-subtracted {\it Chandra}
ACIS-I image in central $200\times200$ kpc$^{2}$($5.8^{\prime}\times5.8^{\prime}$) region
for NGC 6338 group. The VLA 1.4 GHz contours shown in Fig. $5a$ are overlaid.
}
\end{figure}

\begin{figure}
\centering
\includegraphics[width=15.0cm,angle=0]{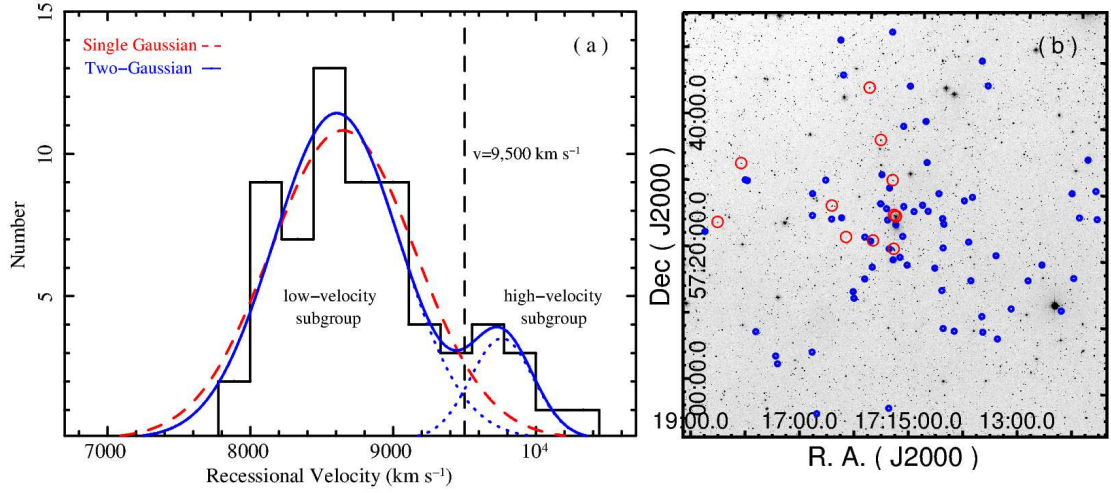}
\caption{
(a) Line-of-sight velocity distribution of 82 member galaxies identified in the NGC 6338 group.
The distribution is fitted with a two-Gaussian model (solid) and a single Gaussian model (dashed; \S4), respectively. (b) DSS optical image
for central $2,100 \times 2,100$ kpc$^{2}$ ($62^{\prime\prime} \times 62^{\prime\prime}$) region of the group,
where the low-velocity ($v<9,500$ km s$^{-1}$) and high-velocity ($v>9,500$ km s$^{-1}$)
member galaxies are marked with small and big circles, respectively.
The positions of 2MASX J1715's two nuclei are very close on the image, which is pointed out with one big circle for clarity.
}
\end{figure}

\end{document}